# Joint Sub-carrier and Power Allocation for Efficient Communication of Cellular UAVs

Hamed Hellaoui, Miloud Bagaa, *Member, IEEE*, Ali Chelli, *Member, IEEE* and Tarik Taleb, *Senior Member, IEEE*

*Abstract*—Cellular networks are expected to be the main communication infrastructure to support the expanding applications of Unmanned Aerial Vehicles (UAVs). As these networks are deployed to serve ground User Equipment (UEs), several issues need to be addressed to enhance cellular UAVs' services. In this paper, we propose a realistic communication model on the downlink, and we show that the Quality of Service (QoS) for the users is affected by the number of interfering BSs and the impact they cause. The joint problem of sub-carrier and power allocation is therefore addressed. Given its complexity, which is known to be NP-hard, we introduce a solution based on game theory. First, we argue that separating between UAVs and UEs in terms of the assigned sub-carriers reduces the interference impact on the users. This is materialized through a matching game. Moreover, in order to boost the partition, we propose a coalitional game that considers the outcome of the first one and enables users to change their coalitions and enhance their QoS. Furthermore, a power optimization solution is introduced, which is considered in the two games. Performance evaluations are conducted, and the obtained results demonstrate the effectiveness of the propositions.

## I. INTRODUCTION

UNMANNED Aerial Vehicles (UAVs) have become an integral part of many applications in different areas (e.g., crowd surveillance, first aid, disaster management, etc.) [2]. The next generation of UAVs will rely on cellular networks as a communication infrastructure. This would allow exploiting the unstopped evolution of these networks and to benefit from their achievements. Cellular-based UAVs would also enable Beyond Visual Line-of-Sight applications (BVLOS), where the UAVs can fly far from their control center. However, in order to enable these potentials, it is highly important to ensure efficient downlink communication of the flying UAVs. Indeed, given their critical nature, any delay in communication could not be tolerable and might have catastrophic consequences.

This work was partially supported by the European Union's Horizon 2020 Research and Innovation Program through the 5G!Drones Project under Grant No. 857031, by the Academy of Finland 6Genesis project under Grant No. 318927, and by the Academy of Finland CSN project under Grant No. 311654.
A preliminary version of this work has been accepted for publication at the IEEE WCNC 2019 [1].
H. Hellaoui and M. Bagaa are with the Department of Communications and Networking, Aalto University, Espoo 02150, Finland (e-mails: hamed.hellaoui@aalto.fi; miloud.bagaa@aalto.fi).
T. Taleb is with the Department of Communications and Networking, Aalto University, Espoo 02150, Finland, also with Centre for Wireless Communications (CWC), University of Oulu, Oulu 90570, Finland, and also with the Department of Computer and Information Security, Sejong University, Seoul 05006, South Korea (e-mail: tarik.taleb@aalto.fi).
A. Chelli is with the Faculty of Engineering and Science, University of Agder, 4898 Grimstad, Norway (e-mail: ali.chelli@uia.no).

This becomes more critical as the flying UAVs can operate in swarm and BVLOS. All these facts highlight the importance of downlink communication in cellular-based UAVs.

To better investigate the communication quality for cellular UAVs, some research works have performed real-field trials [3]–[5]. 3GPP performed relevant evaluations, considering the inputs from several partners. The results provided in [3] point out that flying UAVs could have poor link quality resulting in low throughput. This is mainly due to the close free-space signal propagation, which is translated into a higher probability of interference from non-serving BSs, and thus, leads to poor link quality. This would result in lower Quality of Service (QoS) for the flying UAVs, which may not be tolerable. The effect of this issue on ground UEs is not significant, as the underlying communication model is different compared to that of the flying UAVs. As cellular networks have been so far deployed to serve ground UEs, efficient solutions are highly required for cellular network-enabled UAVs.

### A. Prior Works

Given their potential benefits, the usage of mobile networks for UAVs has attracted a lot of attention. In this context, the quality of the communication link is one of the major issues as the underlying communication model of cellular UAVs is different from that of connected ground UEs [6]. The authors in [7], addressed minimizing the transmit power of BS-UAVs, serving ground users, while satisfying the transmission rate requirements of these users. To this end, the authors investigated the power minimization problem using transport theory and facility location. In [8], authors addressed optimizing 3D placement and mobility of BS-UAVs collecting data from ground IoT. In terms of sub-carriers assignment, the authors used a constrained K-mean clustering strategy to assign different channels to devices that are located in the proximity of each other. As for power control, while the transmit power of each device is computed only based on the channel gain between the device and its serving UAV for the interference-free scenario, an iterative optimization is considered for the interference scenario. The problem of power allocation for multi-UAV enabled wireless networks has been addressed in [9]. The authors proposed a price-based power allocation scheme, and modeled the interaction between the UAVs and the ground users as a Stackelberg game. The ground user selects optimal power strategy to maximize its own utility. The problem of sub-carrier selection and transmission power for UAV-assisted communication is also addressed in [10]. The authors proposed an iterative algorithm and used

Lagrangian dual decomposition method to solve it. In order to enhance the link quality, other works focus on the optimal deployment of BS-UAVs. In [11], the authors addressed the on-demand deployment of multiple aerial base stations for traffic offloading/network recovery. The 3D discrete search space problem is formulated as a mixed-integer program, while a UAV deployment algorithm in a continuous 3D space is proposed based on an unsupervised learning technique. In [12], authors tackled the optimal deployment of multiple UAVs in 3D space to serve ground users. Besides, the paper also considered adapting BS-UAVs' positions as users move around within the network.

In addition to BS-UAVs, some works have addressed the quality of the communication link for cellular-connected drones. Authors in [13] proposed an interference-aware path planing scheme for cellular UAVs. The problem is modeled as a dynamic game among UAVs to achieve a tradeoff between maximizing energy efficiency and minimizing latency and interference level caused on the ground. In [14], a communication model for mobile network-enabled UAVs is proposed. The authors provided an iterative power-optimization algorithm to enhance the link quality on the uplink. In [15], authors considered sub-carrier and power allocation for multi-UAV systems. The problem is addressed by formulating a weighted mean square error (MSE) problem, which can be solved via alternating optimization. In [16], the authors considered the availability of different Mobile Network Operators (MNOs) to steer each UAV's traffic through the network, ensuring the best communication link. The proposed solution is based on the framework of coalitional games in order to associate UAVs to their optimal MNOs. Each UAV will be assigned with sub-carriers only from the selected MNO. In the same context, authors in [17] considered that a UAV could also serve as a relay node to route the traffic of other UAVs and proposed a solution based on linear programming to select the optimal successor for each UAV.

### B. Contributions

This paper studies UAVs' communication in the downlink scenario. We start by considering a communication model that accounts for most of the propagation phenomena experienced by wireless signals. We show that users' QoS depends on the number of interfering BSs and the impact they cause. This leads us to address the joint problem of sub-carrier and power allocation. The major contributions of the paper are the following:

- A realistic communication model for cellular-based UAVs is considered and new expressions for the outage probability are derived. The proposed model accounts for most of the propagation phenomena experienced by wireless signals. Moreover, we show that the quality of service experienced by the UAVs depends on the number of interfering BSs and on the impact they cause.
- We argue and show that separating between the UAVs and the UEs in terms of the assigned sub-carriers would reduce the interference impact on the served users. This is materialized through a matching game, where sub-carriers and users are matched in a way to achieve this objective.
- We propose a coalitional game that builds on the results achieved by the matching game to further reduce the outage probability of each user. This game defines the mechanisms allowing the players to change their coalitions and enhance their QoS. These two games together will constitute two sub-games of a global game.
- A power optimization solution for the communication model is introduced, which is considered in the two sub-games.
- We also provide numerical results on the achieved link quality for performing downlink communication in cellular UAVs.

### C. Organization of the Paper

The rest of the paper is organized as follows. Section II introduces the considered realistic model for cellular network-enabled UAVs. Section III presents the first sub-game optimization based on marching game framework. The second sub-game optimization, based on coalitional game, is introduced in Section IV. Section V provides the global game and the proposed algorithm. Simulations and performance evaluations are conducted and presented in Section VI. Section VII concludes the paper.

## II. SYSTEM MODEL AND PROBLEM FORMULATION

This section describes the proposed communication model and also presents the problem formulation. The *downlink* scenario is considered in which the connected users (UEs and UAVs) receive data from their serving BSs. The latter employs an orthogonal frequency division multiple access (OFDMA) technique to serve the connected users. Consequently, intra-cell interference is neglected, and the interference can be caused only by non-serving BSs, as shown in Fig. 1.

Let $\mathsf{U}$, $\mathsf{V}$ and $\mathsf{B}$ denote respectively the set of UEs, UAVs and BSs. More notations are provided in Table I. Let us also denote by $u$ the serving BS and by $v$ the receiving device. The received signal at the device $v$ considering the sub-carrier $c \in \mathsf{C}$ can be expressed as

$$y_v = \alpha_{uv}\sqrt{P_u}x_u + \sum_{t=1} \alpha_{tv}\sqrt{P_t}x_t + n_v, \quad (1)$$

where $\alpha_{uv}$ refers to the channel gain between the transmitter $u$ and the receiver $v$. The second term in the right-hand side of (1) accounts for the interference impact from non-serving BSs, where $\alpha_{tv}$ is the fading coefficient from the interfering BS $t$ to the receiver $v$. The BSs $u$ and $t$, respectively, transmit the symbols $x_u$ and $x_t$ by employing the power $P_u$ and $P_t$ on the same sub-carrier assigned to the receiving device $v$. As for the third term of (1), $n_v$, it refers to a zero-mean complex additive white Gaussian noise with variance $N_0$. The received signal-to-noise ratio (SNR) for the link $uv$, $\gamma_{uv}$, can be expressed as

$$\gamma_{uv} = P_u|\alpha_{uv}|^2/N_0. \quad (2)$$

As the receiving device $v$ can either be a UE or a UAV, the underlying fading characteristics are substantially different.

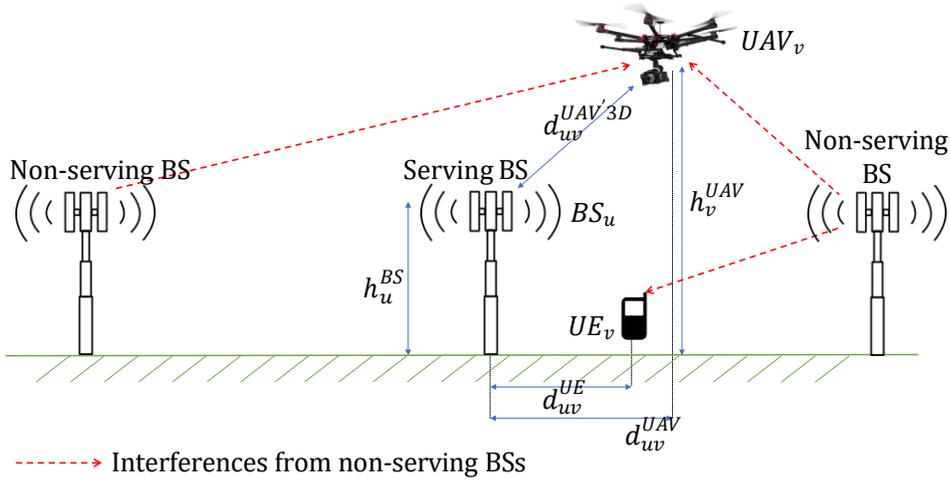

Fig. 1: System model (downlink scenario).

TABLE I: Summary of Notations.

| Notation | Description |
|---|---|
| B | Set of BSs ($|B| = B$). |
| U | Set of UEs ($|U| = U$). |
| V | Set of UAVs ($|V| = V$). |
| A | Set of users $A = U \cup V$ ($|A| = A$). |
| C | Set of sub-carriers ($|C| = C$). |
| $uv$ | Link between the user $v$ and its serving BS $u$. |
| $tv$ | Link between the user $v$ and an interfering BS $t$ ($[1, \ldots, N]$ refer to the list of interfering BSs). |
| $P_{u,c}$ | Power employed by the BS $u \in B$ on the sub-carrier $c \in C$. |
| $P_{uv}$ | minimum transmission power a BS $u$ should employ to send data to the user $v$. |
| $V(u)$ | Set of users connected to the BS $u \in B$. |
| $W(c)$ | Set of users assigned to the sub-carrier $c \in C$. |
| S | Set of coalition; $S = \{S_1, S_2, \ldots, S_C\}$. A coalition $S_c$ is the set of users using the sub-carrier $c \in C$. |
| $w(S)$ | Characteristic function of the coalition S. |
| $\mathbb{I}_S(v)$ | The payoff of the player $v \in S$. |
| $\{S_1, S_2\} d^{v_1} \{S_1', S_2'\}$ | The transfer operation of the user $v_1 \in S_1$ to the coalition $S_2$. The resulting state of the coalitions are respectively denoted by $S_1'$ and $S_2'$. |
| $\{S_1, S_2\} da^{v_1}_{v_2} \{S_1', S_2'\}$ | The operation for exchanging the users $v_1 \in S_1$ and $v_2 \in S_2$. The resulting state of the coalitions are respectively denoted by $S_1'$ and $S_2'$. |

For the case of a UE, the path loss expression provided by 3GPP is considered as [18]

$$PL_{uv}^{UE} = 15.3 + 37.6 \log_{10} d_{uv}^{UE}, \quad (3)$$

where $d_{uv}^{UE}$ refers to the distance in meter between the BS $u$ and the UE $v$ as shown in Fig. 1. A Rayleigh distribution is assumed for the fast fading associated with the UEs. As for the mean SNR of the link between the BS $u$ and the UE $v$, it is denoted by $\bar{\gamma}_{uv}$ and can be expressed as

$$\bar{\gamma}_{uv} = P_u^{UE} \times 10^{-\frac{PL_{uv}^{UE}}{10}} / N_0. \quad (4)$$

The instantaneous received signal-to-interference-plus-noise ratio (SINR) for the link $uv$ can be defined as

$$SINR_{uv} = \frac{\gamma_{uv}}{1 + \sum_{t=1}^{N} \gamma_{tv}}. \quad (5)$$

**Theorem 1.** *Outage probability on the downlink for a UE*

*A UE $v$ fails in receiving packets from its serving BS $u$, on the downlink, iff the $SINR_{uv}$ falls below a threshold $\gamma_{th}$. This event, called outage, occurs with a probability $P_{out,uv}^{UE}$ that can be expressed as*

$$P_{out,uv}^{UE}(\gamma_{th}) = 1 + \exp\left(-\frac{\gamma_{th}}{\bar{\gamma}_{uv}}\right) \sum_{t=1}^{N} \frac{\alpha_t}{\frac{\gamma_{th}}{\bar{\gamma}_{uv}} + \frac{1}{\bar{\gamma}_{tv}}}, \quad (6)$$

*where $\alpha_t$ are unique values satisfying the following equality (fractional decomposition)*

$$\prod_{t=1}^{N}(1 - x\bar{\gamma}_{tv})^{-1} = \sum_{t=1}^{N} \frac{\alpha_t}{x - \frac{1}{\bar{\gamma}_{tv}}}. \quad (7)$$

**Proof:** See Appendix A. □

If the receiving device $v$ is a UAV, the path loss differs depending on the line-of-sight (LoS) and the non-line-of-sight (NLoS) conditions. The LoS communication results in a better QoS compared to the NLoS one. For the proposed solution, we consider the path loss equation provided by 3GPP as [3]

$$PL_{uv}^{UAV} = \begin{cases} 28.0 + 22\log_{10}(d_{uv}^{UAV\;3D}) + 20\log_{10}(f_c) & \text{for LoS link} \\ -17.5 + (46 - 7\log_{10}(h_v^{UAV}))\log_{10}(d_{uv}^{UAV\;3D}) \\ \quad + 20\log_{10}(\frac{40\pi f_c}{3}) & \text{for NLoS link,} \end{cases} \quad (8)$$

where the term $h_v^{UAV}$ refers to the altitude of the UAV $v$ and the term $d_{uv}^{UAV\;3D}$ accounts for the Euclidean distance between the BS $u$ and the UAV $v$, as shown in Fig. 1. The probability of a LoS condition $P_{uv}^{LoS}$ is determined as [3]

$$P_{uv}^{LoS} = \begin{cases} 1 & \text{if } h_v^{UAV} > 100 \\ 1 & \text{if } d_{uv}^{UAV} \leq d_1 \\ \frac{d_1}{d_{uv}^{UAV}} + \exp\left(\frac{-d_{uv}}{p_1}\right)\left(1 - \frac{d_1}{d_{uv}^{UAV}}\right) & \text{if } d_{uv}^{UAV} > d_1. \end{cases} \quad (9)$$

with $p_1 = 4300 \log_{10}(h_v^{UAV}) - 3800$ and $d_1 = \max(460 \log_{10}(h_v^{UAV}) - 700, 18)$. The term $d_{uv}^{UAV}$ refers to the distance between the BS $u$ and the UAV $v$, as illustrated in Fig. 1. It is worth noting that the NLoS probability, $P_{uv}^{NLoS}$, can be obtained as $P_{uv}^{NLoS} = 1 - P_{uv}^{LoS}$. The corresponding fast fading follows a Nakagami-$m$ distribution for LoS links, and a Rayleigh distribution for NLoS links. The mean SNRs of the LoS and NLoS links are denoted by $A_{uv}$ and $B_{uv}$, respectively, and are obtained as

$$A_{uv} = P_{uv}^{LoS} \times P_u/N_0 \times 10^{-\frac{PL_u^{UAV}}{10}} \quad (10)$$
$$B_{uv} = (1 - P_{uv}^{LoS}) \times P_u/N_0 \times 10^{-\frac{PL_u^{UAV}}{10}}.$$

**Theorem 2.** *Outage probability on the downlink for a UAV*

*The outage probability for a UAV $v$ served by the BS $u$ is expressed as*

$$P_{out,uv}^{UAV}(\gamma_{th}) = \sum_{j=1}^{m}\beta_{1j}\frac{(-1)^j}{(j-1)!}\left(\frac{m}{A_{uv}}\right)^{-j}\Gamma(j) + \sum_{t=1}^{N}\alpha'_t f_{j,1}(B_{tv})$$
$$- \sum_{t=1}^{N}\sum_{j'=1}^{m}\frac{\alpha'_{t,j'}(-1)^{j'}}{(j'-1)!}f_{j,j'}(A_{tv}/m) - \beta_{21}B_{uv}\left(1 + \exp(-\frac{\gamma_{th}}{B_{uv}})\right)$$
$$\cdot \sum_{t=1}^{N}\frac{\alpha'_t}{\frac{\gamma_{th}}{B_{uv}} + \frac{1}{B_{tv}}} - \sum_{t=1}^{N}\sum_{j=1}^{m}\frac{\alpha_{t,j}}{\frac{\gamma_{th}}{B_{uv}} + \frac{m}{A_{tv}}}\frac{(-1)^j}{(j-1)!}\Gamma(j) \right), \quad (11)$$

*where $\beta_{1j}$, $\beta_{21}$, $\alpha'_t$ and $\alpha_{t,j}$ are unique values satisfying the two following equations (fractional decomposition):*

$$\left(1 - x\frac{A_{uv}}{m}\right)^{-m}(1 - xB_{uv})^{-1} = \sum_{j=1}^{m}\frac{\beta_{1j}}{(x - \frac{m}{A_{uv}})^j} + \frac{\beta_{21}}{(x - \frac{1}{B_{uv}})} \quad (12)$$

$$\prod_{t=1}^{N}(1-xB_{tv})^{-1}\left(1 - \frac{xA_{tv}}{m}\right)^{-m} = \sum_{t=1}^{N}\frac{\alpha'_t}{x - \frac{1}{B_{tv}}} + \sum_{t=1}^{N}\sum_{j=1}^{m}\frac{\alpha_{t,j}}{x - \frac{m}{A_{tv}}}. \quad (13)$$

*The function $f_{j,j'}(S)$ is provided as*

$$f_{j,j'}(S) = \sum_{p=1}^{n} S^{j'}(\theta_p)^{j'-1}\lambda_p \Gamma\left(j, \frac{m\gamma_{th}(\theta_p S + 1)}{A_{uv}}\right), \quad (14)$$

*where $\lambda_p$ and $\theta_p$ denote the weight and the zero factors of the $n$-th order Laguerre polynomials, respectively [19]. $\Gamma(a,z)$ is the upper incomplete gamma function defined as $\Gamma(a,z) = \int_z^\infty t^{a-1}e^{-t}dt$.*

**Proof:** See Appendix A. □

Theorems 1 and 2 provide the outage probability on the downlink for a ground UE and a UAV, respectively. The outage expressions have been derived by taking into account path loss, fast fading, and interference. This makes the considered system model realistic since it captures most of the propagation phenomena experienced by wireless signals. These expressions are original and can not be found in the literature. This also reflects a major contribution compared to existing works on cellular UAVs as those works do not derive the expression of the outage probability.

The expression of the outage probability in Theorems 1 and 2 do not take the shadowing into account. The reason behind that is the fact that the shadowing is constant during the transmission time of one packet by either a UAV or a UE. The mean power of the signal fluctuates due to shadowing and these fluctuations typically occur on a scale of a few hundred wavelengths [20, p. 69]. Thus the shadowing effect should only be taken into account if the receiver moves a distance of at least 200 wavelengths, which corresponds to a distance of 30 m[1]. In this paper, we focus on the downlink communication between BS and UAVs that is only used for command and control traffic, which is characterized by small packet sizes (maximum of 1250 bytes). Such packets can be transmitted in 0.1 s with a throughput of 100Kbps[2]. In the 3GPP specification [3], it is reported that the maximum UAV speed would not exceed 160 Km/h (44.44 m/s). Thus, during the time required for a packet transmission (0.1 s), the UAV moves a distance of 4.44 m (4.4 m < 30 m). Therefore, the shadowing can be considered constant during a packet transmission from a BS to a UAV. Similarly, for the downlink communication between a BS and UE, it is possible to show that the UE cannot move more than 200 wavelengths (30 m) during a packet transmission, and thus, the shadowing can be considered constant when evaluating the outage for the BS-UE communication. A measurement campaign on multimedia traffic in LTE [21] shows that the maximum packet size, for video call traffic, reaches 1387 Bytes, which is transmitted with an average throughput of 1.21 Mbps. This implies that such a packet has a transmission time of 9.1 ms. Thus, this packet witnesses a constant value of the shadowing because the UE would move less than 0.3 m during the transmission of such a packet if we assume a speed of motion of the UE of 110 km/h (UE is outdoor in a car).

---

[1] The carrier frequency is set to 2GHz; thus, the wavelength equals 0.15 m.

[2] The throughput for command and control traffic has been specified in [3] as equal to 100Kbps.

By carefully analyzing the outage expressions, we can conclude that the outage probabilities depend on the number of interfering BSs and on the impact they are causing as reflected by the terms $\bar{\gamma}_{tv}$ for Theorem 1 and $A_{tv}$ and $B_{tv}$ for Theorem 2. Minimizing the outage probability would be translated into optimizing the sub-carrier assignment and the transmission power on each sub-carrier. These two parameters influence both the number of interfering BSs and their impact. Let $P_c = [P_{1,c}, P_{2,c}, \ldots, P_{B,c}]$ be the transmission power to be used by the $B$ BS for the sub-carrier $c$. We define $P_{out,uv}(P_c)$ as

$$P_{out,uv}(P_c) = \begin{cases} P_{out,uv}^{UAV}(P_c) & \text{if } v \in \mathsf{V} \\ P_{out,uv}^{UE}(P_c) & \text{if } v \in \mathsf{U}, \end{cases} \quad (15)$$

where $P_{out,uv}^{UAV}(P_c)$ and $P_{out,uv}^{UE}(P_c)$ are the outage probabilities for the link $uv$ provided by Theorems 1 and 2 respectively. $P_c$ is the transmission power to be used by the set of BSs on the sub-carrier $c$ assigned to the user $v$. To characterize the selected sub-carrier for the communication between a BS $u$ serving a user $v$, the decision Boolean variable $x_{uv,c}$ is defined as

$$x_{uv,c} = \begin{cases} 1 & \text{If the sub-carrier } c \text{ is used for the link } uv \\ 0 & \text{Otherwise.} \end{cases} \quad (16)$$

Consequently, the joint problem of sub-carrier assignment and power optimization can be expressed as

$$\min \max_{v \in \mathsf{A}} P_{out,uv}(P_c) \quad (17)$$

**s.t.**

$$\forall u \in \mathsf{B}, \forall v \in V(u); \sum_{c \in \mathsf{C}} x_{uv,c} = 1 \quad (18)$$

$$\forall u \in \mathsf{B}, \forall c \in \mathsf{C}; \sum_{v \in V(u)} x_{uv,c} \leq 1 \quad (19)$$

$$\forall u \in \mathsf{B}, \forall c \in \mathsf{C}; 0 \leq P_{u,c} \leq P_{max}. \quad (20)$$

The objective function in (17) is to minimize the outage probability for the the set of users connected to the cellular network. Both sub-carrier assignment and power optimization are considered for minimizing the outage probability. (18) ensures that each user $v$ is assigned with one sub-carrier from its serving BS $u$. Here, $V(u)$ refers to the set of users connected to the BS $u$, as mentioned in Table I. (19) captures the fact that a sub-carrier is used at most by one user. (20) is the feasibility constraint for the transmission power. The formulated optimization problem is a non-linear program, which is complex to solve, especially for large networks. This complexity is inherent from the proposed communication model, which considers most of the propagation phenomena characterizing wireless communication.

As discussed previously, the QoS on the downlink is affected by the number of interfering BSs and the impact they cause. For this purpose, the joint problem of sub-carrier assignment with power optimization is considered. We propose a solution based on the framework of game theory. The game is divided into two independent sub-games: a matching sub-game and a coalitional sub-game. The matching sub-game, $\mathsf{G}_1$ is executed at first. It aims at matching the users to the sub-carriers in a way to meet the preferences of each other. The preferences are defined in a way to optimize the outage probability of the users and to enhance the downlink communication of cellular-based UAVs. The second sub-game, $\mathsf{G}_2$ builds on the outcome assignment of the matching game to boost this optimization. It is based on the framework of coalitional games and defines operations to enable users to change their coalitions in order to enhance their QoS. The power optimization is taken into account in the two sub-games. The following sections explain our proposed solution in more detail.

### III. MATCHING SUB-GAME OPTIMIZATION

The framework of many-to-one matching game [22] is adopted to model the problem. A related projection of this framework is the college admissions problem where colleges are receiving the applications of students and each college has a fixed quota for accommodation. Both colleges and students have an order of preference. The goal is to model the interactions between these entities and to achieve a matching stability satisfying as much as possible their preferences.

The game $\mathsf{G}_1$ is defined as $\mathsf{G}_1 = (\mathsf{A}, \mathsf{C}, \geq^v, \geq_c)$, where $\mathsf{A} = \mathsf{U} \cup \mathsf{V}$ is the set of UEs and UAVs (considered as students) while $\mathsf{C}$ is the set sub-carriers (considered as colleges). $\geq^v$ and $\geq_c$ are, respectively, the preference relation for a user $v \in \mathsf{A}$ and a sub-carrier $c \in \mathsf{C}$. The goal of the game is to associate each user $v \in \mathsf{A}$ to a sub-carrier $c \in \mathsf{C}$ while meeting certain constraints and preferences. Indeed, each sub-carrier $c$ has a quota of only $B$ (number of BSs). Moreover, a sub-carrier can accommodate at most one user connected to a given BS. This is due to the fact that each of the connected users to a given BS will be served by different sub-carriers. Let $W(c)$ be the set of users assigned to the sub-carrier $c$. Consequently, the matching game should maintain the following condition:

$$\forall c \in \mathsf{C}, \forall u \in \mathsf{U}; \; |W(c)^T V(u)| \leq 1. \quad (21)$$

To define the order of preference, one would rely on equations (6) and (11) as they express the objective function to minimize (minimizing the outage probability for each user). However, as it can be seen from those equations, computing the outage probability for a user depends on the terms $\bar{\gamma}_{tv}$ (for a UE) or $A_{tv}$ and $B_{tv}$ (for a UAV). These terms reflect the interference part from the non-serving BSs $t$ using the same sub-carrier as $v$. In other words, an interdependence is present between users to specify a sub-carrier, based on (6) and (11), as it depends on the choice of the other users in using the same sub-carrier or not. This makes the direct usage of those equations for defining the order of preference very complex.

In order to set a preference order for the users, we propose to rely on minimizing the interference impact over the link $tv$ for each user $v$. We first evaluate the interference amount over the links $tv$. The associated terms reflect the mean SNR from the interfering BS to the user $v$ ($\bar{\gamma}_{tv}$ in the case of a UE, and $A_{tv}$ and $B_{tv}$ in the case of a UAV). The formulas for the mean SNR are expressed in (4) and (10). Therefore, the terms related to the links $tv$ depend on the power employed by the interfering BS for transmitting data to the served users. The power value would be larger than the one employed to serve users in an interference-free situation. Indeed, if a user fails in receiving packets from its serving BS in an interference-free environment, it will definitely fail in the presence of interference. We consider this fact to estimate the minimum transmission power employed by a BS $u$ to serve its user $v$.

**Lemma 1.** *A UE $v$ will fail in receiving data from its serving BS $u$ iff the transmission power employed by the latter is less than a value given as*

$$\bar{P}_{uv} = SNR_{th} N_0 / 10^{-\frac{PL_{uv}^{UE}}{10}}, \quad (22)$$

*where, $SNR_{th}$ reflects the receiver sensitivity. As for a UAV $v$, the value is given as*

$$\bar{P}_{uv} = SNR_{th} N_0 / (P_{uv}^{LoS} 10^{-\frac{PL_{uv}^{UAV}}{10}} + P_{uv}^{NLoS} 10^{-\frac{PL_{uv}^{UAV}}{10}}). \quad (23)$$

Equations (22) and (23) of the above lemma can be derived from equations (4) and (10), respectively. They provide the minimum transmission power a BS $u$ should employ to send data to the user $v$. Below this value, the user will definitely fail in receiving the data. A numerical evaluation of $\bar{P}_{uv}$ for both UEs and UAVs, considering the same $SNR_{th}$, shows that UAVs require less transmission power compared to ground UEs. This result is due to the better channel condition characterizing aerial communication. Consequently, the interference impact, over the link $tv$, is bigger when the interfering BS $t$ is serving a UE compared to the case where it is serving a UAV. We exploit this fact to order the preferences in a way to reduce the interference impact. The following two rules are introduced.

1) As per the previous discussion, users of the same type (either UEs or UAVs) would prefer to be gathered in the same sub-carriers. Indeed, the heterogeneity in terms of channel conditions between flying UAVs and ground UEs is behind the issue of cellular UAVs. Due to the close free-space signal propagation, a flying UAV might require less power to receive data on the downlink than what is required for a UE on the ground having the same distance. Consequently, having UEs and UAVs on the same sub-carrier might result in considerable interference on UAVs from the BSs serving UEs. Separating between UAVs and UEs in the assigned sub-carriers would reduce this impact. This constitutes the preference of the users.
2) The sub-carriers will receive the application of the users. The admitted users within a given sub-carrier will be subject to interference from non-serving BSs whose served users are also admitted in this sub-carrier. As discussed before, the interference terms reflect the mean SNR and depend on the power employed by the non-serving BSs. To reduce the interference impact, each sub-carrier would prefer having users requiring less transmission power from their serving BSs. In addition, a sub-carrier accommodating users requiring large transmission powers would admit fewer users. The power value $\bar{P}_{uv}$ is used to compare between users' transmission power. This represents the basis for sub-carriers to order their preferences of the users.

Considering the first rule, the users are not able to know which sub-carriers they are gathered in, so they can make their choices and set the preferences. In order to overcome this issue, we consider that the sub-carriers are ordered $\mathfrak{C} = [c_1 \ldots c_C]$). Consequently, users from the same type can easily be gathered in the same sub-carriers by choosing to be near to one of the two ends. We consider that the UEs would prefer being in the first sub-carriers while the UAVs would rather be interested in the last ones. As for the second rule, a sub-carrier uses the transmission power $\bar{P}_{uv}$ to order its preferences from the candidate users of each BS. It prefers at most one user per BS, thus maintaining condition (21).

Based on this discussion, we define the preference relation for users/sub-carriers and the admission criteria as follows.

**Definition 1.** *Ordering of preferences and admission criteria*

*For a UE $v \in \mathbb{U}$, the preference order of the sub-carriers is defined as follows:*

$$Pref^{UE}(v) \iff c_1 \geq^v c_2 \geq^v \cdots \geq^v c_C. \quad (24)$$

*For a UAV $v \in \mathbb{V}$ the preference order of the sub-carriers is defined as follows*

$$Pref^{UAV}(v) \iff c_C \geq^v \cdots \geq^v c_2 \geq^v c_1. \quad (25)$$

*Each sub-carrier $c \in \mathfrak{C}$ receives applications of the candidate users and prefers at most one user per BS. The preference relation $\geq_c$ is defined between two users $v_1$ and $v_2$ connected to the same BS $u$ ($v_1, v_2 \in V(u)$) as follows*

$$v_1 \geq_c v_2 \iff \bar{P}_{uv_1} < \bar{P}_{uv_2}. \quad (26)$$

The above definition allows ordering the preferences for both users and sub-carriers. As discussed before, UEs prefer being in the first sub-carriers while the UAVs would rather be interested in the last ones. This is, respectively, reflected by equations (24) and (25) of Definition 1. Here, $c_i \geq^v c_j$ means that the user $v$ prefers the sub-carrier $c_i$ on $c_j$. On the other hand, the sub-carriers use the transmission power $\bar{P}_{uv}$ to establish the preference order of the candidate users connected to a BS $u$. As previously discussed, the preferred user is the one having the smallest value of $\bar{P}_{uv}$. This is materialized through equation (26) of Definition 1.

The matching will result in assigning the UEs to the first sub-carriers and the UAVs to the last ones. When there are some BSs having less users than the number of sub-carriers, the matching will end up with some free sub-carriers in between the two types of users. However, this means that a sub-carrier in the middle might have no users while others are filled. In order to overcome this situation, we introduce virtual users in the network in a way to complete the network. These virtual users are considered to complete the number of users for each BS. They will be removed once the matching is done. Moreover, in order to enable sub-carriers to accommodate users requiring large transmission powers to admit fewer users (rule 2), the virtual users are distributed to have more large values of the $\bar{P}_{uv}$ parameter. A half-normal distribution is used to distribute $\bar{P}_{uv}$ between the smallest and the biggest values.

One important notion in matching game is the stability. It defines the situation where players are matched and have no incentive to change their association. The matching algorithm should lead to a stable situation, which is defined as follows.

**Definition 2.** *Matching stability*

*The matching is said to be unstable if there are two users $v_1$ and $v_2$ connected to the same BS $u$ ($v_1, v_2 \in V(u)$) such as $v_1$ is matched to $c_1$ and $v_2$ is matched to $c_2$, although $v_2$ prefers $c_1$ to $c_2$ and $c_1$ prefers $v_2$ to $v_1$. Formally, this is expressed as*

$$\exists u \in S, \exists v_1, v_2 \in V(u), \exists c_1, c_2 \in C;$$
$$v_1 \in W(c_1) \text{ and } v_2 \in W(c_2) \text{ and } c_2 \geq^{v_1} c_1 \text{ and } v_1 \geq_{c_2} v_2. \quad (27)$$

Once the sub-carriers are matched with the users, a power optimization procedure is considered. This would allow reducing the interference impact from non-serving BSs, after the matching process, while maintaining the outage probability for the concerned users under a certain threshold. The power optimization procedure will be considered for each sub-carrier of the set of BSs. As it can be noticed form (6) and (11), the outage probability for a user is impacted by the transmission power employed by the serving and interfering BSs. This is referred respectively by the terms $\bar{\gamma}_{uv}$ and $\bar{\gamma}_{tv}$ in Theorem 1 and by the terms $A_{uv}, B_{uv}, A_{tv}$ and $B_{tv}$ in Theorem 2. This shows the interdependence between the transmission powers that should be employed by the different BSs, and also the complexity of deriving these powers from equations (6) and (11). On the other hand, when the transmission powers of the interfering BSs are unchanged (over the links $tv$), the outage probability of the user $v$ decreases as the transmission power from the serving BS increases (over the link $uv$). This can be derived from the outage probability definition, which is $P_{out}(\gamma_{th}) = P(\gamma_{uv}/(1 + \sum_{t=1}^{N} \gamma_{tv}) \leq \gamma_{th})$. This consideration is exploited to propose the power optimization described through Algorithm 1.

Algorithm 1 presents the power optimization procedure for a sub-carrier $c \in C$ of all the BSs. Initially, the maximum transmission power is used (line 1). For each BS $u \in B$, if it is serving a user $v$ through the sub-carrier $c$ (lines 5 and 6), a power optimization is considered; The transmission power of the BS $u$ is reduced until having the maximum outage probability smaller than a threshold $P_{out,th}$, while maintaining the power larger than the minimum value, $\bar{P}_{uv}$ (lines [8-10]). As explained in the previous paragraph, when the transmission powers from the interfering BSs are unchanged, the outage probability for the user $v$ increases as the transmission power from the serving BS decreases. Since changing the transmission power of a BS also influences the outage probability of users subject to interference, this power optimization is repeated. Lines [11-14] evaluate if the transmission power has changed after executing lines [8-10]. The variable 'Stable' is used to characterize the stable situation, where no BS needs to reduce its transmission power. This will allow reducing the transmission power of the BSs while maintaining desired outage probabilities.

The parameters $P^{max}$, $\bar{P}_{uv}$ and $P^{stp}$ denote, respectively, the maximum value for the transmission power, the minimum value for the transmission power, and the unit value for reducing the power. The complexity of Algorithm 1 in the worst case scenario is $O(\dot{S} * B * (\frac{P^{max} - \bar{P}_{uv}}{P^{stp}} + 1))$, where $\dot{S}$ reflects the number of unstable solutions found before the effective one. Indeed, in the worst case scenario, lines [8-10] will be executed $\frac{P^{max} - \bar{P}_{uv}}{P^{stp}} + 1$ times for each BS $u \in B$. It is difficult to determine the value of $\dot{S}$, as it depends on the outage probability expressions, reflected in Theorems 1 and 2, which are very complex. We, therefore, provide in the performance evaluation section some obtained results for $\dot{S}$ along with the considered parameters. In the other hand, $O((\frac{P^{max} - \bar{P}_{uv}}{P^{stp}} + 1)^B)$ can be considered as the upper bound of the complexity (at most, reducing the power will be evaluated $\frac{P^{max} - \bar{P}_{uv}}{P^{stp}} + 1$ times at each BS).

---

**Algorithm 1** Power optimization for users using $c \in C$.

**Input:** $P_{out,th}, SNR_{th}, P^{stp}, P^{max}$
1: $P_c = P^{max}$
2: **while** True **do**
3:    Stable = True
4:    **for** each BS $u \in S$ **do**
5:       **if** $W(c) \cap V(u) \mathrel{/}= null$ **then**
6:          $v = W(c) \cap V(u)$
7:          $\acute{P}_c = P_c$
8:          **while** $P_{out,uv}\ [P_{1,c}, P_{2,c}, \ldots, P_{u,c} - P^{stp}, \ldots, P_{B,c}] \leq P_{out,th}$ **and** $P_{u,c} - P^{stp} > \bar{P}_{uv}$ **do**
9:             $P_c = [P_{1,c}, P_{2,c}, \ldots, P_{u,c} - P^{stp}, \ldots, P_{B,c}]$
10:          **end while**
11:          **if** $\acute{P}_c \neq P_c$ **then**
12:             $\acute{P}_c = P_c$
13:             Stable = False
14:          **end if**
15:       **end if**
16:    **end for**
17:    **if** Stable **then**
18:       Break
19:    **end if**
20: **end while**
21: **return** $\acute{P}_c$

---

## IV. COALITIONAL SUB-GAME OPTIMIZATION

The matching game provided in the previous section will result in sub-carrier assignment with power optimization. This assignment is a heuristic and might not be optimal. The objective of the second sub-game is to boost this optimization by progressively reducing the outage probability for each user. To this end, we consider the framework of coalitional game. The cooperative nature of this game would allow the enforcement of the optimization achieved in the first sub-game.

Formally, the game $G_2$ is defined as $G_2 = (A, S, w)$; with $A$ being the set of users (UEs and UAVs) and $S$ the set of coalitions. A coalition $S_c$ represents the set of users using the sub-carrier $c \in C$. The initial coalition assignment is indeed the result of the first matching game. Having said that, the number of coalitions is exactly the number of sub-carriers. We can therefore write $S = \{S_1, S_2, \ldots, S_C\}$ with $S_c \subseteq A$. We note that each two coalitions involves entirely two different set of players; i.e. $\forall S_1, S_2 \in S : S_1 \cap S_2 = \emptyset \Rightarrow S_1 \cap S_2 = \emptyset$ (a user connected to a BS will be served using only one sub-carrier). $w$ represents the characteristic function and is defined by the payoff of each player, $w(S_c) = (\Pi_{S_c}(v), v \in S_c)$, as

$$\Pi_{S_c}(v) = 1 - P_{out,uv}(\hat{P}_c). \quad (28)$$

As it can be seen from (28), the outage probability is considered in the definition of the payoff of the corresponding player belonging to a coalition. This is coupled with power optimization, $\hat{P}_c$, leading therefore to an optimized outage probability for all the players in the coalition. As for the benefit of a coalition, it is defined as

$$w(S_c) = \sum_{v \in S_c} (\Pi_{S_c}(v)). \quad (29)$$

In the context of coalitional game, the users change from one coalition to another in order to have better payoff. Within our framework, we define two types of operations to enable users to change their coalitions and increase their benefits. These operations are user transfer and user exchange. In the first case, a user belonging to a coalition will be transferred to another one. This could happen when the receiving coalition does not have another user connected to the same BS as the candidate user. In the second case, two users connected to the same BS will be exchanged from their respective coalitions. The execution of these operations would result in two new coalitions instead of the originals[3]. The transfer or the exchange would be useful if it leads to enhanced payoff for the players belonging to the new resulting coalitions. Formally, we define the transfer and the exchange rules as follows:

**Definition 3.** *Transfer and exchange rules*

A user $v_1 \in S_1$ served by a BS $u$ would be transferred to another coalition $S_2$ (does not have another user served by the same BS; i.e., $V(u) \cap S_2 = \emptyset$), resulting respectively into two coalitions $S_1'$ and $S_2'$, iff:

$$\{S_1, S_2\} d^{v_1} \{S_1', S_2'\} \Leftrightarrow w(S_1') - w(S_1) > w(S_2) - w(S_2'). \quad (30)$$

Two users $v_1 \in S_1$ and $v_2 \in S_2$ served by the same BS $u$ $(v_1, v_2 \in V(u))$ would be exchanged, resulting into two coalitions $S_1'$ and $S_2'$ respectively, iff:

$$\{S_1, S_2\} d^{v_1}_{v_2} \{S_1', S_2'\} \Leftrightarrow \begin{cases} \Pi_{S_2'}(v_1) \geq \Pi_{S_1}(v_1) \text{ and} \\ \Pi_{S_1'}(v_2) \geq \Pi_{S_2}(v_2) \end{cases} \quad (31.1)$$

$$\text{And} \begin{cases} \forall v \in S_1 \cap S_1' : \Pi_{S_1'}(v) \geq \Pi_{S_1}(v) \text{ and} \\ \forall v \in S_2 \cap S_2' : \Pi_{S_2'}(v) \geq \Pi_{S_2}(v) \end{cases} \quad (31.2)$$

$$\text{And} \begin{cases} \Pi_{S_2'}(v_1) > \Pi_{S_1}(v_1) \text{ or} \\ \Pi_{S_1'}(v_2) > \Pi_{S_2}(v_2) \text{ or} \\ \exists v \in S_1 \cap S_1' : \Pi_{S_1'}(v) > \Pi_{S_1}(v) \text{ or} \\ \exists v \in S_2 \cap S_2' : \Pi_{S_2'}(v) > \Pi_{S_2}(v). \end{cases} \quad (31.3)$$

$$(31)$$

The definition in (30) means that a player $v_1$ will be transferred from the coalition $S_1$ to the coalition $S_2$, resulting respectively into two coalitions $S_1'$ and $S_2'$, if the gain of this operation on the first coalition is larger than the loss on the second coalition. Indeed, transferring a player from a coalition (without exchange) would be translated into withdrawing an interferer from the coalition. This would enhance the outage probability of the coalition's users and increase the underlying utility function. On the other hand, the receiving coalition will have a new user causing interference and leading therefore to decreased utility function. The transfer operation would be approved only when the gain is greater than the loss. As for the exchange rule, the definition in (31) means that two players $v_1 \in S_1$ and $v_2 \in S_2$ will be exchanged, and leading to two coalitions $S_1'$ and $S_2'$, if at least the payoff of one player will be increased after the operation (through conditions in (31.3)), while that of all other players remain unaffected (including both the exchanged players (31.1) and the other players (31.2)). We note that while the transfer rule focus on the global benefit and payoff of the coalitions, the exchange rule is more centered on the individual profit of the users.

The users change from one coalition to another as defined in the above definition. This process is repeated iteratively. The situation (partition of coalitions) where there is no incentive to do a change is said to be stable.

**Definition 4.** *Coalitions stability*

A partition of coalitions is said to be stable when the transfer and exchange rules can not be applied (defined in equations (30) and (31)), i.e.,

$$\begin{cases} \forall S_1, S_2 \in S, \forall v_1 \in S_1; \quad \nexists S_1', S_2' \subseteq A : \{S_1, S_2\} d^{v_1} \{S_1', S_2'\} \\ \text{and} \\ \forall S_1, S_2 \in S, \forall v_1 \in S_1, v_2 \in S_2; \nexists S_1', S_2' \subseteq A : \{S_1, S_2\} d^{v_1}_{v_2} \{S_1', S_2'\} \end{cases} \quad (32)$$

## V. GLOBAL GAME AND GENERAL ALGORITHM

Based on the framework of game theory, we propose two sub-games to enhance communication in cellular UAVs. While the matching sub-game provides an initial assignment of users to sub-carriers with power optimization, the coalitional game boosts this optimization. Algorithm 2 shows the execution steps of the global game.

---

[3] The number of coalitions is always the same. We mean by 'new coalitions' the original ones after executing the operations (transfer or exchange).

**Algorithm 2** Global game.
**Input:** $A = C \cup U$, $S$
// 1st sub-game
1: **while** $A \setminus \bigcup_{c \in C} W(c) \neq \emptyset$ **do**
2: $\quad T = A \setminus \bigcup_{c \in C} W(c)$
3: $\quad Q^{UE} = T \cap U$
4: $\quad Q^{UAV} = T \cap U$
   // UEs apply for their preferred sub-carriers using $Pref^{UE}(v)$
5: $\quad$ **if** $Q^{UE} \neq \emptyset$ **then**
6: $\quad\quad c$ = lower sub-carrier not yet targeted by UEs
7: $\quad\quad D = \{v \in Q^{UE}, v \succeq_c v' \; \forall v' \in Q^{UE}, V(u)\}$
8: $\quad\quad W(c) = W(c) \cup D$
9: $\quad$ **end if**
   // UAVs apply for their preferred sub-carriers using $Pref^{UAV}(v)$
10: $\quad$ **if** $Q^{UAV} \neq \emptyset$ **then**
11: $\quad\quad c$ = upper sub-carrier not yet targeted by UAVs
12: $\quad\quad D = \{v \in Q^{UAV}, v \succeq_c v' \; \forall v' \in Q^{UAV}, V(u)\}$
13: $\quad\quad W(c) = W(c) \cup D$
14: $\quad$ **end if**
   // Rejected users will apply for their next preferences
15: **end while**
16: $P = \hat{P}$ // for each $c \in C$ using Algorithm 1
   // 2nd sub-game
17: Initialize $E$ from $C$
18: **while** True **do**
19: $\quad$ Stable = True
20: $\quad$ **for** each two coalitions $E_1, E_2$ **do**
21: $\quad\quad$ **for** each BS $u \in S$ **do**
22: $\quad\quad\quad$ **if** $V(u)^T E_1 \neq \emptyset$ and $V(u)^T E_2 \neq \emptyset$ **then**
23: $\quad\quad\quad\quad$ let $v1 = V(u)^T E_1$ and $v2 = V(u)^T E_2$
        // try exchanging $v1$ and $v2$
24: $\quad\quad\quad\quad$ **if** $\{E_1, E_2\} \lhd_{v2}^{v1} \{E_1', E_2'\}$ **then**
25: $\quad\quad\quad\quad\quad$ $E_1 = E_1'$; $E_2 = E_2'$
26: $\quad\quad\quad\quad\quad$ Stable = False
27: $\quad\quad\quad\quad$ **end if**
28: $\quad\quad\quad$ **else if** $V(u)^T E_1 \neq \emptyset$ **then**
29: $\quad\quad\quad\quad$ let $v1 = V(u)^T E_1$
        // try the transfer of $v1$
30: $\quad\quad\quad\quad$ **if** $\{E_1, E_2\} \lhd^{v1} \{E_1', E_2'\}$ **then**
31: $\quad\quad\quad\quad\quad$ $E_1 = E_1'$; $E_2 = E_2'$
32: $\quad\quad\quad\quad\quad$ Stable = False
33: $\quad\quad\quad\quad$ **end if**
34: $\quad\quad\quad$ **else if** $V(u)^T E_2 \neq \emptyset$ **then**
35: $\quad\quad\quad\quad$ let $v2 = V(u)^T E_2$
        // try the transfer of $v2$
36: $\quad\quad\quad\quad$ **if** $\{E_2, E_1\} \lhd^{v2} \{E_1', E_2'\}$ **then**
37: $\quad\quad\quad\quad\quad$ $E_1 = E_1'$; $E_2 = E_2'$
38: $\quad\quad\quad\quad\quad$ Stable = False
39: $\quad\quad\quad\quad$ **end if**
40: $\quad\quad\quad$ **end if**
41: $\quad\quad$ **end for**
42: $\quad$ **end for**
43: $\quad$ **if** Stable **then**
44: $\quad\quad$ Break
45: $\quad$ **end if**
46: **end while**

The matching sub-game will assign a sub-carrier $c \in C$ to each user $u \in A$ (line 1). This assignment respects the ordering of preferences provided in Definition 1. In the logic of the assignment, users (UEs or UAVs) apply for their preferred sub-carriers, while the latter accept or reject the candidates according to their preferences. Lines 3 and 4 identify, respectively, the UEs and the UAVs not yet assigned to a sub-carrier. As per the preference order of the users, the most preferred sub-carrier for the UEs is the lower one while the upper one is the most preferred sub-carrier for the UAVs (equations (24) and (25) respectively). This is materialized in lines 6 and 11 respectively. The target sub-carriers accept the users as per their preference order defined in equation (26). The admission of UEs and UAVs is shown in lines [7-8] and lines [12-13], respectively. Each rejected user will apply for its next preferred sub-carrier.

The coalitional sub-game considers the outcome of the matching game as initial partition of coalitions (line 17). Thereafter, the transfer and exchange operations will be attempted. For each two coalitions $S_1$ and $S_2$ and each BS $u \in B$ (lines [20-21]), the operations will be attempted depending on whether the BS has players in these coalitions. If the BS has a player in each coalition, an exchange operation will be attempted (lines [22-27]). However, if the BS has a player in only one coalition, the transfer operation will be attempted to the other one (lines [28-33] and lines [34-39]). As defined in Definition 3, the transfer and exchange operations will be approved only when they lead to enhanced payoff. The variable '*Stable*' is used to characterize the stability situation as defined in Definition 4; i.e., the situation where no exchange or transfer rules need to be applied.

As shown in Algorithm 2, the execution of the global game starts with the matching sub-game, then the coalitional one. The convergence of the global game depends on the convergence of the two sub-games. In what follows, we prove the convergence of the proposed matching and coalitional sub-games.

**Theorem 3.** *Matching game convergence*

*The matching game of Algorithm 2 (1st sub-game) is guaranteed to converge to a stable matching.*

*Proof:* From Definition 2, a matching would be unstable if a user and a sub-carrier consider the change an improvement. This is materialized by equation (27). Let us recall the instability situation;

$$\exists u \in S, \exists v_1, v_2 \in V(u), \exists c_1, c_2 \in C; \underbrace{v_1 \in W(c_1) \text{ and } v_2 \in W(c_2)}_{(33.a)} \text{ and } \underbrace{c_2 \succeq_{v_1} c_1}_{(33.b)} \text{ and } \underbrace{v_1 \succeq_{c_2} v_2}_{(33.c)}, \quad (33)$$

where the part (33.a) corresponds to a matching achieved after executing the algorithm, while the parts (33.b) and (33.c) reflect the instability. Let us suppose that equation (33) is correct after the execution of the first sub-game of Algorithm 2. According to this Algorithm, the users connected to each BS apply for the sub-carriers according to their order of preferences. This is defined by equation (24) for the UEs and (25) for the UAVs.

At most, two sub-carriers will receive the candidates as the UEs prefer the first sub-carriers while the UAVs prefer the last ones. Each sub-carrier prefers the users having the smallest value of $\bar{P}_{uv}$ from each BS, as defined in (26). The users that got rejected will apply for their next preferred sub-carriers. Having said that, the UEs of each BS will be in the first sub-carriers while the UAVs will be in the last ones. Moreover, the UEs are ordered in the sub-carriers according to their $\bar{P}_{uv}$ values while the UAVs according to the inverse of this parameter.

For two sub-carriers $c_1$ and $c_2$, the order can be either $c_1$ before $c_2$ or the inverse. Let us suppose that $c_1$ is ordered before $c_2$ (the same logic holds if $c_2$ has an order before $c_1$) and consider the part (33.a) of equation (33), which we are supposing its correctness. As per the previous paragraph (the order of the users in the sub-carriers), three situations are possible for $v_1$ and $v_2$, which are the following

$$(33.a) \Rightarrow \begin{cases} v_1 \in U \text{ and } v_2 \in V & (34.1) \\ \text{Or} \\ v_1, v_2 \in U \text{ and } \bar{P}_{uv_1} \leq \bar{P}_{uv_2} & (34.2) \\ \text{Or} \\ v_1, v_2 \in V \text{ and } \bar{P}_{uv_1} \geq \bar{P}_{uv_2} & (34.3) \end{cases} \quad (34)$$

If we consider the two situations (34.1) and (34.2), $v_1 \in U$, they are in contradiction with the part (33.b), $c_2 \geq^{v_1} c_1$. Indeed, if $c_1$ is ordered before $c_2$, UEs would prefer $c_1$ on $c_2$ (equation (24)). Now, the only situation that might hold is (34.3). As $c_1$ is ordered before $c_2$, and $v_1$ has bigger value of the $\bar{P}_{uv}$ parameter than $v_2$ (UAVs are ordered according the inverse of the parameter $\bar{P}_{uv}$, as per the previous discussion). Nevertheless, this also contradicts the part (33.c), $v_1 \geq^{c_2} v_2$, as a sub-carrier prefers users with smaller value of the parameter $\bar{P}_{uv}$. Consequently, regarding equation (33), if (33.a) holds, (33.b) or (33.c) do not. This proves the incorrectness of the instability assumption and concludes the proof of Theorem 3. □

As per the above proof, the matching sub-game is stable and converges to a final partition. We provide in the following the convergence proof of the coalitional sub-game, which implies, together with the matching stability, the convergence of the global game.

**Theorem 4.** *Coalitional game convergence*

*Starting from an initial partition of the players on the coalitions $S = \{S_1, S_2, \ldots, S_C\}$ the coalitional game of Algorithm 2 (2nd sub-game) is guaranteed to converge to a final and optimal partition.*

*Proof:* The initial partition of the players on the coalitions $S$ is the result of the first sub-game. This coalition will be the subject of transformations applied sequentially. Let us express this sequence by the following equation

$$S^{(0)} = S^{(G_1)} \rightarrow S^{(1)} \rightarrow S^{(2)} \rightarrow S^{(3)} \ldots \quad (35)$$

where $S^{(0)}$ is the initial state of the coalitions which is the result of the matching game $S^{(G_1)}$. $\rightarrow$ reflects the applied transformation which can be a transfer or an exchange operation. $S^{(i)}$ is the state of the coalitions after the $i^{th}$ transformation. As the number of coalitions, the number of players and the number of states a player can have in a coalition (used transmission power) are limited, the number of partitions is also limited. We therefore define the following lemma.

**Lemma 2.** *The convergence of the sequence of equation (35) is guaranteed when the transformations do not lead to repeated partitions.*

The above lemma is based on the fact that the number of partitions is limited. If the produced partitions are not repeated, this would lead to a final partition $S^{(final)}$. In addition, as each derived partition is optimized compared to the previous (transfer and exchange are approved only when there is enhancements), the final partition is optimal. Consequently, proving Theorem 4 comes down to proving that the derived partitions do not repeat. To this end, we consider two types of operations: transfer and exchange. In the case of a transfer operation, equation (30) is considered. This equation can also be written as

$$\{S_1, S_2\} d^{v_1} \{S_1', S_2'\} \Leftrightarrow w(S_1') + w(S_2') > w(S_2) + w(S_1) \quad (36)$$

which implies that the resulting coalitions, together, have better payoff than the originals. In addition, user transfer from one coalition to another does not affect the other coalitions (i.e., $\forall S_i \in S \setminus \{S_1, S_2\}, w(S_i)$ remains the same). Consequently, we can write the following

$$S^{(i)} \rightarrow S^{(i+1)} \Leftrightarrow \sum_{S \in S^{(i)}} w(S) < \sum_{S \in S^{(i+1)}} w(S). \quad (37)$$

On the other hand, the exchange operation is ruled by equations (31). As it is clearly stated by these equations, the exchange is approved only when at least the payoff of one player is increased while those of the other players, in the two coalitions, remain unchanged. Besides, as in the case of transfer, the other coalitions are not affected by the exchange operation. These facts guarantee the equivalence in equation (37) even for the exchange operation. This means that each derived partition is different from the previous and therefore not repeated. Moreover, the sum of benefits of the coalitions increases after each transformation. This concludes the proof of Theorem 4. □

Algorithm 2 reflects the global game which is composed of two sub-games with different complexities. The first sub-game (lines [1 - 15]) represents the matching game, and its complexity is $O(A)$ (matching each user to a sub-carrier). This complexity can be further reduced to $O(\frac{A}{2*B})$ if the UEs and the UAVs at each BS $u \in B$ are being matched in parallel to the sub-carriers, as presented in Algorithm 2. As for the second sub-game (lines [18 - 46]), it reflects the coalitional sub-game and its complexity in the worst-case scenario is $O(\ddot{S} * C(C-1) * B)$, where $\ddot{S}$ is the number of unstable solutions found before reaching the effective one. The term $C(C-1)$ refers to the number of possible combinations of two coalitions, while $B$ is the number of BSs. Indeed, the transfer/exchange operation will be evaluated between every two coalitions for users belonging to the same BS $u \in B$. It is very difficult to determine the value of $\ddot{S}$ as it depends on the outage probabilities, which are very complex. Therefore, we provide in the performance evaluation section a discussion on the results associated with $\ddot{S}$ and the considered parameters. On the other hand, $O((C(C-1))^B)$ can be considered as the upper bound of the complexity (at most, $C(C-1)$ combinations will be evaluated at each BS). We can see that the first sub-game is associated with a lower complexity compared to the second one. While the first game is meant to provide a quick solution reflected in a heuristic, the second game starts with the achieved partition and boosts the optimization to further reduce the outage probabilities.

Therefore, we will also compare the complexity of the second game with the brute-force search solution.

**Execution of the Algorithm**

Algorithm 2 intends to be executed in a centralized way by the cellular network providing communication to the users. It aims to optimize the allocation of power and sub-carriers to the connected users in a way to achieve enhanced QoS. The latter is represented by the outage probabilities for UEs and UAVs. Executing this Algorithm requires information on the base stations and also on the connected users. On the users' side, the geographical position of each UE and UAV is required. On the base stations' side, their geographical positions are required in addition to the used sub-carriers and transmission powers that will be fixed by the Algorithm. All this information is mainly provided by the cellular network. Indeed, the latter owns the base stations and has access to their geographical positions, their transmission powers and their used sub-carriers. Furthermore, the geographical position of the users can be known in advance (e.g., prior to its flight, the UAV operator submits the flight plan to the UTM -Unmanned aerial system Traffic Management- which can be communicated to the mobile operator) or be provided by the mobile network using MPS (Mobile Positioning System). All this information will feed the Algorithm to allow the cellular network to provide enhanced QoS to the served users. On the other hand, the proposed solution allows optimizing the outage probability for the connected users without considering their mobility (or using prior knowledge on their mobility). Extending the current solution to cover the mobility of the users will be addressed as future work.

## VI. Performance Evaluation

This section provides the performance evaluation of the proposed solution for efficient communication in cellular-based UAVs. The solution aims to enhance the QoS for the UAVs, reflected by the outage probability, by jointly optimizing the sub-carrier and power allocation. The simulation parameters are summarized in Table II.

Fig. 2 illustrates the average outage probability of UAVs and UEs for different values of the parameter $\gamma_{th}$. These results were obtained by using 50 UEs and 50 UAVs in the simulation. Fig. 2 shows that the outage increases as the value of $\gamma_{th}$ increases. The parameter $\gamma_{th}$ indicates the threshold that the SINR should exceed in order to have successful reception of the packets. It represents the receiver sensitivity threshold. As $\gamma_{th}$ increases, the receiver's ability to detect weak signals decreases. This highlights the importance of choosing a good value for $\gamma_{th}$. In the rest of the evaluations, the value $10^{-1}$ is considered for the parameter $\gamma_{th}$. Moreover, we can also see from Fig. 2 that on average, the outage probability for the UAVs is larger than that of the UEs. This is mainly due to the better channel condition characterizing aerial communication, which is translated into more interference on the UAVs. This complies with the results of real-field evaluations and shows the issue of downlink communication in cellular UAVs.

TABLE II: Summary of simulation parameters.

| Parameter | Value |
|---|---|
| Simulation area | 1000m x 1000m |
| Altitude of the UAVs | 22.5 - 300 m [3] |
| BSs | 10 BSs (each with 11 sub-carriers) |
| $P^{max}$, $P^{stp}$ | 20w, 0.1w (resp.) |
| $f_c$ | 2 GHz |
| $N_0$ | -130 $dBm$ [20] |
| Nakagami parameter $m$ | 2 |
| $P_{out,th}$ | 0.1 |

To further investigate this problem, we illustrate in Fig. 3 the impact of the number of devices (UEs and UAV) on the average outage probability. We notice that for both UEs and UAVs, the outage probability increases as the number of deployed devices increases. However, the outage probability of UAVs increases faster compared to the outage probability of UEs. This shows that as the number of devices in the network increases, it becomes challenging to maintain a good link quality for UAVs as they are more sensitive to interference. Therefore, efficient solutions for downlink communication of UAVs are highly required.

In Fig. 4, we evaluate the performance of our proposed solution in reducing the average outage probability of UAVs. Additionally, we compare our proposed algorithm to two baseline solutions. The first baseline solution uses random assignment of the sub-carriers, while the second baseline solution is inspired from graph theory and utilizes maximum independent set for sub-carrier assignment. The second baseline solution was proposed in [1]. The average outage probability is evaluated for different values of the number of devices in the network, and the obtained results are presented in Fig. 4. The line labeled 'power optimization' refers to the case when directly performing the power optimization on the random partition without considering the rules for assigning the sub-carriers in the matching sub-game. This allows us to verify the validity of the proposed rules. By comparing this result with that from considering the first sub-game, we can see that the proposed rules show their effectiveness, especially when the network is saturated. From this figure, we also notice that our proposed solution allows us to reduce the average outage probability significantly compared to the other baseline solutions. Moreover, the gain of using our solution becomes more significant as the number of devices in the network increases. In this way, if a predefined QoS is required for the set of UAVs, our proposed solution allows us to accommodate more devices in the network compared to the baseline solutions. Compared to the random partition, the first sub-game allowed to reduce the outage probability by 13.72% and 35.11% when considering 70 and 100 users in the network, respectively. As for the second sub-game, it allowed reducing the outage probability by 57.62% and 48.71% when considering 70 and 100 users in the network, respectively. We have also evaluated the number of unstable solutions for power optimization before the effective one. This reflects the parameter $S^*$ considered in the definition of the complexity of Algorithm 1 as $O(S^* \cdot B_* \cdot (\frac{P^{max}-P_{uv}}{P_{stp}} + 1))$. Through the different evaluations, $S^*$ was on average equal to 6,42. It has also reached 14 as a maximum value. This result complements the evaluation of the complexity of Algorithm 1 provided in Section III.

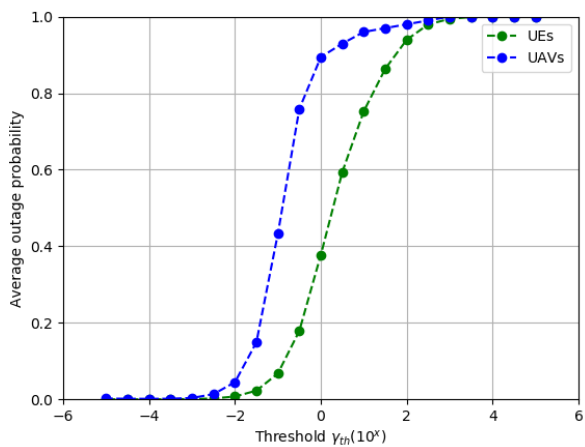

Fig. 2: Outage probabilities $P_{out,uv}(\gamma_{th})$ for different threshold values $\gamma_{th}$.

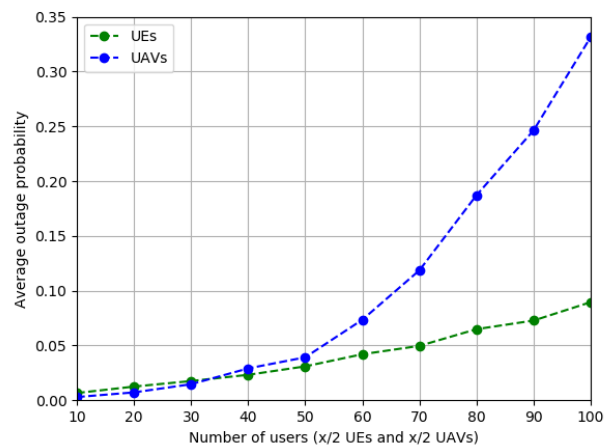

Fig. 3: Effect of increasing the number of UEs and UAVs on the outage probability.

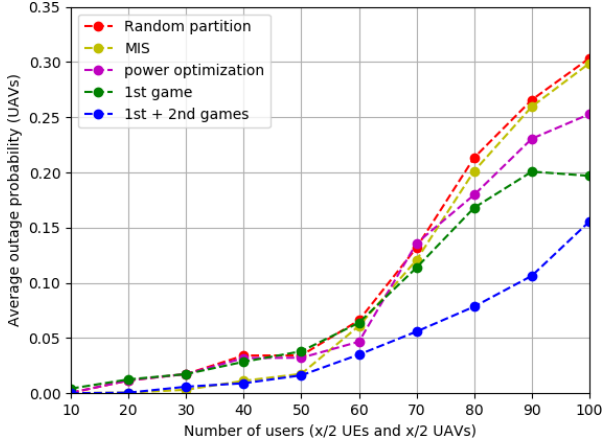

Fig. 4: Evaluation of the outage probability for the proposed solutions.

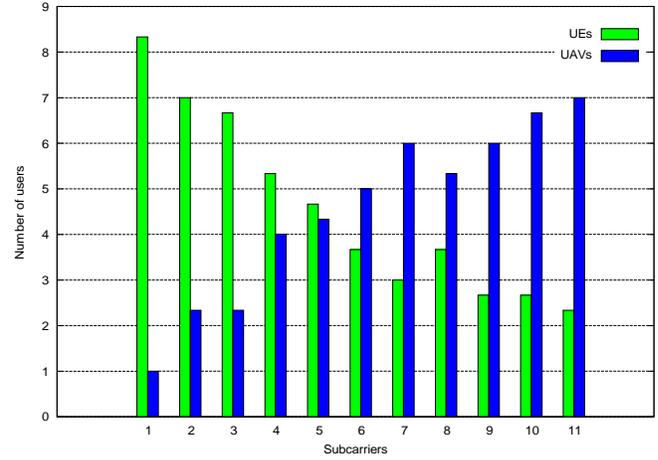

Fig. 5: User distribution on the sub-carriers after running the game.

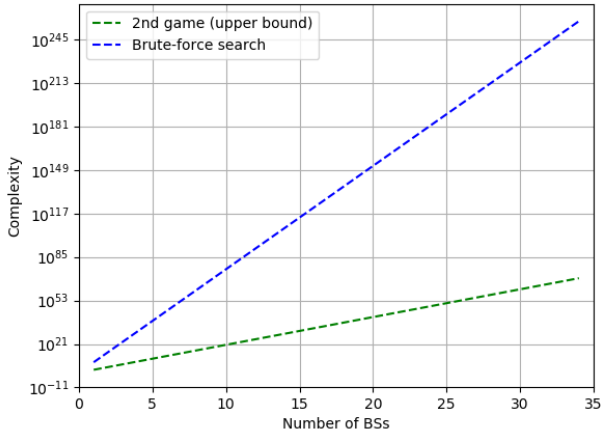

Fig. 6: Comparison between the complexity of brute-force search with the coalitional game

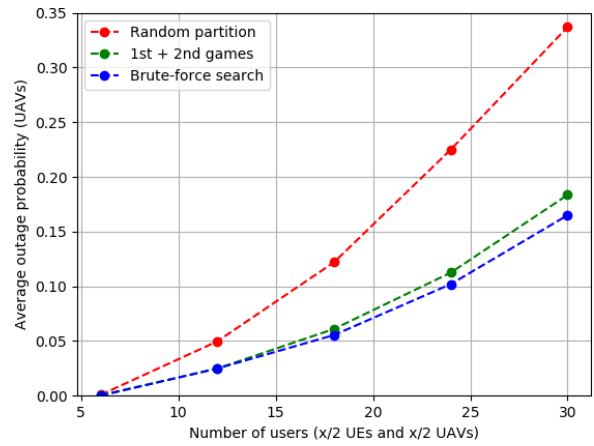

Fig. 7: Comparison of the proposed solution with brute-force search (10 BSs).

Fig. 5 provides the distribution of users on the sub-carriers after executing Algorithm 2 (Global game). As detailed in the previous section, the last sub-game is based on the framework of coalitional game. It allows users to change their coalitions in a way to enhance their payoffs. The obtained results show that, after reaching stability, more UEs are assigned to the first sub-carriers while more UAVs are assigned to the last ones. This comes to support the idea defended in this paper that separating between UAVs and UEs in terms of the assigned sub-carriers would enhance the quality of communication. The heterogeneity of the channel conditions between the two types of users is translated into different power requirements to be employed by the serving BS. Consequently, having different types of users in the same sub-carriers could increase the interference impact.

As mentioned in Section V the matching sub-game is very quick and its complexity is $O(A)$. This complexity can even be reduced to $O(\frac{A}{2*B})$ when users belonging to different BSs are assigned in parallel. The second sub-game is meant to further reduce the outage probabilities by performing a series of transfer/exchange operations starting from the partition achieved by the first one. The complexity of the second sub-game is $O(\ddot{S} * C(C-1) * B)$, where $\ddot{S}$ is the number of unstable solutions found before reaching the effective one. Through the different evaluations, the average value of $\ddot{S}$ was 4.36 and has reached 6 as a maximum value. We have compared the complexity of the second sub-game (upper bound, $O((C(C-1))^B)$) with that of the brute-force search solution. The complexity of the latter is $O((C!)^B)$. Indeed, $C!$ corresponds to the number of combinations of the users on the sub-carriers of a given BS. The evaluation is reflected in Fig. 6 (each BS has 11 users each assigned to a sub-carrier). As it can be seen from this figure, the coalitional game is associated with a very low complexity compared to the brute-force search. Furthermore, we have also implemented the algorithm that looks for the optimal solution (brute-force search) and compared the achieved average outages probabilities with that from the proposed solution. The obtained results are shown in Fig. 7. This evaluation is considered for a small network of 10 BSs, where each one has 3 sub-carriers. As we can see, the proposed solution achieves a near-optimal solution that has the same trend as the optimal solution generated by the brute-force search, especially when the network is not complete (some sub-carriers are free).

## VII. CONCLUSION

In this paper, we addressed the issue of downlink communication in cellular network-enabled UAVs. Given the lack of realistic communication models that reflect the underlying phenomena experienced by wireless signals when communicating with flying UAVs, we introduced a model that accounts for path loss, fast fading, and interference. The analysis of this model showed that the quality of service experienced by the UAVs is affected by the number of interfering BSs and the impact they cause. We, therefore, considered the joint problem of sub-carrier assignment and power optimization. Leveraging the framework of game theory, we proposed two sub-games to tackle this complex problem. This complexity is inherent from the underlying realistic model. The first sub-game is based on matching game theory. It implements the proposition that separating between the ground UEs and the flying UAVs in terms of the assigned sub-carriers would reduce the interference impact and enhance the QoS. The second sub-game is based on coalitional game. It organizes the players based on the outcome of the first game and defines the mechanisms allowing them to change their coalitions and to enhance the users' QoS. Furthermore, a power optimization algorithm is proposed for this communication model, which is considered in the two sub-games. The performed evaluations demonstrated the effectiveness of the proposed solution in enhancing communication in cellular UAVs.

## APPENDIX A
## APPENDIX: PROOF OF THEOREMS 1 AND 2

This appendix provides the proofs of Theorems 1 and 2. The outage probability expressions on the downlink are derived for both a UE and a UAV. This probability is defined as $P_{out}(\gamma_{th}) = P(SINR \leq \gamma_{th})$. Let us recall the expression of the SINR which is given as

$$SINR_{uv} = \frac{\gamma_{uv}}{1 + \sum_{t=1}^{N} \gamma_{tv}} = \frac{\gamma_{uv}}{1 + I'} = \frac{\gamma_{uv}}{I'} \quad (A.1)$$

where, the term $I' = \sum_{t=1}^{N} \gamma_{tv}$ includes all the interfering BSs. Consequently, the outage probability for the link $uv$ can be written as

$$P_{out}(\gamma_{th}) = P(SINR \leq \gamma_{th}) = P\left(\frac{\gamma_{uv}}{I'} \leq \gamma_{th}\right)$$
$$= E_{I'}\left[P(\gamma_{uv} \leq \gamma_{th}y | I' = y)\right] = \int_{0}^{\infty} F_{\gamma_{uv}}(\gamma_{th}y) P_{I'}(y) dy \quad (A.2)$$

where the term $F_{\gamma_{uv}}(x)$ refers to the Cumulative Distribution Function (CDF) of $\gamma_{uv}$ (computed as $F_{\gamma_{uv}}(x) = \int_{0}^{x} P_{\gamma_{uv}}(y) dy$), while $P_{I'}(y)$ is the Probability Density Function (PDF) of $I'$. The expression of these functions differs on whether the receiver equipment $v$ is a UE or a UAV.

**The receiver equipment $v$ is a UE**
Let us compute the Moment Generating Function (MGF) and the PDF of $\gamma_{uv}$. Their expressions can be given as

$$M_{\gamma_{uv}}^{UE}(s) = (1 - s\bar{\gamma}_{uv})^{-1} \quad (A.3)$$

$$P_{\gamma_{uv}}^{UE}(x) = \frac{1}{\bar{\gamma}_{uv}} \exp\left(-\frac{x}{\bar{\gamma}_{uv}}\right). \quad (A.4)$$

The MGF of $I'$ includes all the interfering BSs and can be deduced as

$$M_{I'}(s) = \prod_{t=1}^{N} M_{\gamma_{tv}}(s) = \prod_{t=1}^{N}(1 - s\bar{\gamma}_{tv})^{-1} = \sum_{t=1}^{N} \frac{\alpha_t}{\bar{s} - \frac{1}{\bar{\gamma}_{tv}}} \quad (A.5)$$

where $\alpha_t$ is the same as in Theorem 1, satisfying (7), and is obtained using fractional decomposition (multinomial theorem [23]). The PDF $P_{I'}(x)$ of $I'$ can be obtained, by computing the inverse Laplace transform of $M_{I'}(s)$ in (A.5), as

$$P_{I'}(x) = L^{-1}[M_{I'}(s)] = L^{-1}\left[\sum_{t=1}^{N} \frac{\alpha_t}{s - \frac{1}{\bar{\gamma}_{tv}}}\right] = \sum_{t=1}^{N} \alpha_t(-1)\exp\left(-\frac{x}{\bar{\gamma}_{tv}}\right). \quad (A.6)$$

Using (A.6) and the fundamental theorem of transformation of random variables [24], the PDF $P_{I'}(y)$ of $I'$ is computed as

$$P_{I'}(y) = \sum_{t=1}^{N} \alpha_t'(-1) \exp\left(-\frac{y-1}{\bar{\gamma}_{tv}}\right). \quad (A.7)$$

As for the CDF $F_{\gamma_{uv}}^{UE}(x)$ of $\gamma_{uv}$, it is determined from (A.4)

$$F_{\gamma_{uv}}^{UE}(x) = \int_0^x P_{\gamma_{uv}}^{UE}(y)dy = 1 - \exp\left(-\frac{x}{\bar{\gamma}_{uv}}\right). \quad (A.8)$$

Consequently, the outage probability can be computed as

$$P_{out}^{UE}(\gamma_{th}) = \int_1^\infty F_{\gamma_{uv}}(\gamma_{th}y) P_{I'}(y) dy$$
$$= 1 - \int_1^\infty \exp\left(-\frac{\gamma_{th}y}{\bar{\gamma}_{uv}}\right) P_{I'}(y) dy$$
$$= 1 + \exp\left(-\frac{\gamma_{th}}{\bar{\gamma}_{uv}}\right) \sum_{t=1}^{} \frac{\alpha_t'}{\frac{\gamma_{th}}{\bar{\gamma}_{uv}} + \frac{1}{\bar{\gamma}_{tv}}}. \quad (A.9)$$

Note that (A.9) is the outage probability provided in Theorem 1. □

**The receiver equipment $v$ is a UAV**

Aerial communication is characterized by both LoS and NLoS conditions. As defined in (10), the terms $A_{uv}$ and $B_{uv}$ reflect the mean SNR related to the two conditions. Let us compute the MGF $M_{\gamma_{uv}}^{UAV}(s)$ of $\gamma_{uv}$.

$$M_{\gamma_{uv}}^{UAV}(s) = M_{\gamma_{LoS,uv}} M_{\gamma_{NLoS,uv}} = \left(1 - \frac{sA_{uv}}{m}\right)^{-m}(1-sB_{uv})^{-1}$$
$$= \sum_{j=1}^{} \frac{\beta_{1j}}{(s-\frac{m}{A_{uv}})^j} + \frac{\beta_{21}}{(s-\frac{1}{B_{uv}})} \quad (A.10)$$

where $\beta_{1j}$ and $\beta_{21}$ are the same as in Theorem 2, satisfying (12). They are obtained using fractional decomposition. The PDF of $P_{\gamma_{uv}}^{UAV}(x)$ can be obtained from the MGF $M_{\gamma_{uv}}^{UAV}(s)$ of $\gamma_{uv}$ using the inverse Laplace transform as

$$P_{\gamma_{uv}}^{UAV}(x) = \sum_{j=1}^{m} \left[\beta_{1j} L^{-1}\left\{\frac{1}{(s-\frac{m}{A_{uv}})^j}\right\} + \beta_{21} L^{-1}\left\{\frac{1}{(s-\frac{1}{B_{uv}})}\right\}\right]$$
$$= \sum_{j=1}^{} \beta_{1j} x^{j-1} \exp\left(-\frac{mx}{A_{uv}}\right) \frac{(-1)^j}{(j-1)!} - \beta_{21} \exp\left(-\frac{x}{B_{uv}}\right). \quad (A.11)$$

As for the MGF of $I$ it is computed as

$$M_I(s) = \prod_{t=1}^{} M_{\gamma_{tv}}(s) = \prod_{t=1}^{} (1-sB_{tv})^{-1} \left(1-\frac{sA_{tv}}{m}\right)^{-m}$$
$$= \sum_{t=1}^{} \frac{\alpha_t'}{s-\frac{1}{B_{tv}}} + \sum_{t=1}\sum_{j=1} \frac{\alpha_{t,j}}{(s-\frac{m}{A_{tv}})^j} \quad (A.12)$$

where $\alpha_t'$ and $\alpha_{t,j}$ are the same as in Theorem 2, satisfying (13). They are obtained using fractional decomposition. Now, we can compute the PDF of $I$ as

$$P_I(x) = L^{-1}[M_I(s)] = L^{-1}\left[\sum_{t=1}^{N} \frac{\alpha_t'}{s-\frac{1}{B_{tv}}} + \sum_{t=1}\sum_{j=1} \frac{\alpha_{t,j}}{(s-\frac{m}{A_{tv}})^j}\right]$$
$$= \sum_{t=1}^{N} -\alpha_t' \exp\left(-\frac{x}{B_{tv}}\right) + \sum_{t=1}\sum_{j=1} \alpha_{t,j} \frac{(-1)^j x^{j-1}}{(j-1)!} \exp\left(-\frac{mx}{A_{tv}}\right). \quad (A.13)$$

Considering the fundamental theorem of transformation of random variables, $P_{I'}$ is obtained as

$$P_{I'}(y) = \sum_{t=1}^{} \alpha_t'(-1) \exp\left(-\frac{y-1}{B_{tv}}\right)$$
$$+ \sum_{t=1}\sum_{j=1} \alpha_{t,j} \frac{(-1)^j}{(j-1)!} \exp\left(-\frac{m(y-1)}{A_{tv}}\right)(y-1)^{j-1}. \quad (A.14)$$

As for the CDF of $\gamma_{uv}$, it is computed as

$$F_{\gamma_{uv}}^{UAV}(x) = \int_0^x P_{\gamma_{uv}}^{UAV}(y)dy = \sum_{j=1}^{} \beta_{1j} \frac{(-1)^j}{(j-1)!} \underbrace{\int_0^x y^{j-1}\exp\left(-\frac{my}{A_{uv}}\right)dy}_{K_1} - \beta_{21}\underbrace{\int_0^x \exp\left(-\frac{y}{B_{uv}}\right)dy}_{K_2} \quad (A.15)$$

with

$$K_1 = \left(\frac{m}{A_{uv}}\right)^{-j}\left[\Gamma(j) - \Gamma\left(j, \frac{mx}{A_{uv}}\right)\right]$$
$$K_2 = B_{uv}\left[1 - \exp\left(-\frac{x}{B_{uv}}\right)\right]. \quad (A.16)$$

Finally, the outage probability can be computed as

$$P_{out}(\gamma_{th}) = \int_1^\infty F_{\gamma_{uv}}(\gamma_{th}y) P_{I'}(y)dy = \sum_{j=1}^{} \beta_{1j} \frac{(-1)^j}{(j-1)!}\left(\frac{m}{A_{uv}}\right)^{-j}\left[\Gamma(j) - \int_1^\infty \Gamma\left(j, \frac{m\gamma_{th}y}{A_{uv}}\right) P_{I'}(y)dy\right] - \beta_{21} B_{uv}\left[1 - \int_1^\infty \exp\left(-\frac{\gamma_{th}y}{B_{uv}}\right)P_{I'}(y)dy\right] \quad (A.17)$$

$$= \sum_{j=1}^{} \beta_{1j}\frac{(-1)^j}{(j-1)!}\left(\frac{m}{A_{uv}}\right)^{-j}\left[\Gamma(j) + \sum_{t=1}\alpha_t'\sum_{p=1}^{n} B_{tv}\lambda_p \Gamma\left(j, \frac{m\gamma_{th}(\theta_p B_{tv}+1)}{A_{uv}}\right) - \sum_{t=1}\sum_{j'=1}\alpha_{t,j'}\frac{(-1)^{j'}}{(j'-1)!}\sum_{p=1}^{n}(A_{tv}/m)^{j'}\lambda_p\theta_p^{j'-1}\Gamma\left(j, \frac{m\gamma_{th}(\theta_p(A_{tv}/m)+1)}{A_{uv}}\right)\right] - \beta_{21}B_{uv}\left[1 + \sum_{t=1}\alpha_t'\frac{\exp(-\frac{\gamma_{th}}{B_{uv}})}{\frac{\gamma_{th}}{B_{uv}}+\frac{1}{B_{tv}}} - \sum_{t=1}\sum_{j=1}\alpha_{t,j}\frac{(-1)^j}{(j-1)!}\left(\frac{\gamma_{th}}{B_{uv}}+\frac{m}{A_{tv}}\right)^{-j}\exp\left(-\frac{\gamma_{th}}{B_{uv}}\right)\Gamma(j)\right] \quad (A.18)$$

$$= \sum_{j=1}^{} \beta_{1j}\frac{(-1)^j}{(j-1)!}\left(\frac{m}{A_{uv}}\right)^{-j}\left[\Gamma(j) + \sum_{t=1}^{N}\alpha_t' f_{j,1}(B_{tv}) - \sum_{t=1}\sum_{j'=1}\alpha_{t,j'}\frac{(-1)^{j'}}{(j'-1)!}f_{j,j'}(A_{tv}/m)\right] - \beta_{21}B_{uv}\left[1+\exp\left(-\frac{\gamma_{th}}{B_{uv}}\right)\sum_{t=1}^{N}\frac{\alpha_t'}{\frac{\gamma_{th}}{B_{uv}}+\frac{1}{B_{tv}}} - \sum_{t=1}\sum_{j=1}^{m}\frac{\alpha_{t,j}}{\left(\frac{\gamma_{th}}{B_{uv}}+\frac{m}{A_{tv}}\right)^j}\frac{(-1)^j}{(j-1)!}\Gamma(j)\right]. \quad (A.19)$$

The integrals in (A.17) involve the Gamma function with the exponential function. The Laguerre polynomial, defined as $\int_0^\infty e^{-x}f(x)dx = \sum_{p=1}^{n}\lambda_p f(\theta_p)$, is used to perform a numerical evaluation, where $\lambda_p$ and $\theta_p$ are the weight and the zero factors of the $n$-th order Laguerre polynomials, respectively. The result in (A.18) is obtained using a change of variable. Note that the expression of $f_{j,j'}(S)$ is provided in equation (14). The result in (A.19) is the same as the outage probability expression presented in Theorem 2. □

Note that Theorem 1 is a special case of Theorem 2. Using the expression of the outage in (A.19), we can derive the following 4 special cases.

**Special Case 1**: $P_{uv}^{LoS} = 0$, $P_{uv}^{NLoS} = 1$, $P_{tv}^{LoS} = 0$, and $P_{tv}^{NLoS} = 1$. This implies that in (A.10) the parameters $\beta_{1j} = 0$ and $\beta_{21} = 1/B_{uv}$, whereas in (A.13) the parameters $\alpha_{t,j} = 0$ and $\alpha_t' = 1/B_{tv}$. Thus, the outage probability for special case 1 can be computed using (A.19) as

$$P_{out}(\gamma_{th}) = 1 + \exp\left(-\frac{\gamma_{th}}{B_{uv}}\right) \sum_{t=1} \frac{\alpha_t'}{\frac{\gamma_{th}}{B_{uv}} + \frac{1}{B_{tv}}}, \quad (A.20)$$

which matches with the outage probability provided in Theorem 1.

**Special Case 2**: $P_{uv}^{LoS} = 1$, $P_{uv}^{NLoS} = 0$, $P_{tv}^{LoS} = 1$, and $P_{tv}^{NLoS} = 0$. This implies that in (A.10) the parameters $\beta_{1j} \neq 0$ and $\beta_{21} = 0$, whereas in (A.13) the parameters $\alpha_{t,j} \neq 0$ and $\alpha_t' = 0$. Thus, the outage probability for special case 2 can be computed using (A.19) as

$$P_{out}(\gamma_{th}) = \sum_{j=1} \beta_{1j} \frac{(-1)^j}{(j-1)!} \left(\frac{m}{A_{uv}}\right)^{-j}$$
$$\left(\Gamma(j) - \sum_{t=1}\sum_{j'=1}^{m} \alpha_{t,j'} \frac{(-1)^{j'}}{(j'-1)!} f_{j,j'}(A_{tv}/m)\right). \quad (A.21)$$

**Special Case 3**: $P_{uv}^{LoS} = 1$, $P_{uv}^{NLoS} = 0$, $P_{tv}^{LoS} = 1$, $P_{tv}^{NLoS} = 0$, and $m = 1$. This implies that in (A.10) the parameters $\beta_{11} = 1/A_{uv}$, $\beta_{1j} = 0$ for $j = 2, \ldots, m$ and $\beta_{21} = 0$, whereas in (A.13) the parameters $\alpha_{t,1} = 1/A_{tv}$, $\alpha_{t,j} = 0$ for $j = 2, \ldots, m$, and $\alpha_t' = 0$. Thus, the outage probability for special case 3 can be computed using (A.19) as

$$P_{out}(\gamma_{th}) = 1 + \exp\left(-\frac{\gamma_{th}}{A_{uv}}\right) \sum_{t=1} \frac{\alpha_{t,1}'}{\frac{\gamma_{th}}{A_{uv}} + \frac{1}{A_{tv}}}. \quad (A.22)$$

**Special Case 4**: $P_{uv}^{LoS} = 0$, $P_{uv}^{NLoS} = 1$, $P_{tv}^{LoS} \neq 0$, and $P_{tv}^{NLoS} \neq 0$. This implies that in (A.10) the parameters $\beta_{1j} = 0$ and $\beta_{21} = 1/B_{uv}$, whereas in (A.13) the parameters $\alpha_{t,j} \neq 0$ and $\alpha_t' = 1/B_{tv}$. Thus, the outage probability for special case 4 can be computed using (A.19) as

$$P_{out}(\gamma_{th}) = 1 + \exp\left(-\frac{\gamma_{th}}{B_{uv}}\right)\left(\sum_{t=1} \frac{\alpha_t'}{\frac{\gamma_{th}}{B_{uv}} + \frac{1}{B_{tv}}}\right.$$
$$\left. - \sum_{t=1}^{N}\sum_{j=1}^{m} \frac{\alpha_{t,j}}{\frac{\gamma_{th}}{B_{uv}} + \frac{m}{A_{tv}}} \frac{(-1)^j}{(j-1)!} \Gamma(j)\right). \quad (A.23)$$